\documentclass[aps,preprint,showpacs]{revtex4}
\usepackage{graphicx}

\def\al{\alpha}
\def\be{\beta}

\def\si{\sigma}

\def\De{\Delta}

\def\bfi{{\mathbf{i}}}
\def\bj{{\mathbf{j}}}

\def\br{{\mathbf{r}}}

\def\bx{{\mathbf{x}}}

\newcommand{\ben}{\begin{equation}}
\newcommand{\een}{\end{equation}}
\newcommand{\bea}{\begin{eqnarray}}
\newcommand{\eea}{\end{eqnarray}}
\newcommand{\ba}{\begin{array}}
\newcommand{\ea}{\end{array}}
\newcommand{\bi}{\begin{itemize}}
\newcommand{\ei}{\end{itemize}}
\def\math{\mathsurround 0pt}
\def\oversim#1#2{\lower.5pt\vbox{\baselineskip0pt \lineskip-.5pt

\ialign{$\math#1\hfil##\hfil$\crcr#2\crcr{\scriptstyle\sim}\crcr}}}

\def\pa{\partial}
\def\half{\frac{1}{2}}

\newcommand{\vev}[1]{\langle#1\rangle}
\newcommand{\A}{A}

\begin{document}

\preprint{SUSX-TH/02-029, arXiv:hep-ph/0212350}
\date{December 2002}
\title{Scaling in Numerical Simulations of Domain Walls}

\author{Theodore Garagounis}
\email[]{t.garagounis@sussex.ac.uk}
\author{Mark Hindmarsh}
\email[]{m.b.hindmarsh@sussex.ac.uk}

\affiliation{Centre for Theoretical Physics \\
 University of Sussex \\
 Brighton BN1 9QJ \\
 U.K}

\date{\today}

\begin{abstract}
We study the evolution of domain wall networks appearing after
phase transitions in the early Universe. They exhibit interesting
dynamical scaling behaviour which is not yet well understood, and
are also simple models for the more phenomenologically acceptable
string networks. We have run numerical simulations in two- and
three-dimensional lattices of sizes up to $4096^3$.
The theoretically predicted scaling solution for
the wall area density $ A $ $ \propto 1/t $
is supported by the simulation results, while no evidence of a
logarithmic correction reported in previous studies could be
found. The energy loss mechanism appears to be direct 
radiation, rather than the formation and collapse of closed loops or 
spheres.  We discuss the implications for the evolution of 
string networks.
\end{abstract}


\maketitle


\section{Introduction}

The idea that the symmetries in nature are not respected by the
vacuum plays a crucial role in the unification of forces.
Moreover, any broken symmetries that can be identified today were
likely to have been restored at high temperatures
\cite{Kirzhnits:1972iw,Dolan:1974qd,Kirzhnits:1976ts}. These two facts suggest
that the Universe underwent phase transitions in its early
history. It was realised that phase transitions may leave the
Universe filled with topological defects
\cite{Zeldovich:1974uw,Kibble:1976sj}, which if massive enough
could be observed through their density fluctuations
\cite{Kibble:1976sj,Zeldovich:1980gh,Vilenkin:1983jv} (see also
\cite{VilShe94,Hindmarsh:1995re} for reviews).

Probably the most interesting and important property of defect
networks is that they seem to exhibit dynamic scaling. This means
that they quickly lose memory of their initial conditions and
evolve towards configurations which can be characterised by a
single length scale (or perhaps a few \cite{Austin:1995dz}) $ \xi
$. This length scale is thought to increases with time with a
universal exponent. In a relativistic field theory one can argue
from dimensional analysis that $ \xi (t)\sim tf(Mt) $, where $ M
$ is the mass scale of the defect. The large-scale dynamics of
defects are independent of $M$, and so the dynamical scaling
exponent can be naively estimated as 1.

This behaviour has been checked by numerically simulating
classical field theories for domain walls
\cite{Press:1989yh,Coulson:1996nv,Larsson:1997sp}, gauge strings
\cite{Vincent:1998cx,Moore:1998gp,Moore:2001px}, global strings
\cite{Ryden:1989vj,Yamaguchi:2000dy}, 
global monopoles \cite{Bennett:1993fy,Yamaguchi:2001rf,Yamaguchi:2001xn}
and textures 
\cite{Bennett:1993fy,Durrer:1994da,Pen:1994nx}. All the
simulations are consistent with the linear scaling law over the
range of the simulations, but do allow other behaviours: in
particular, Press, Ryden and Spergel suggested that the results
for domain walls would be better fitted by $ \xi \sim t/\ln (t) $.

Since the original naive scaling arguments were put forward, a more 
quantitative
approach to the dynamics of domain wall networks has been proposed by one of
the authors \cite{Hindmarsh:1996xv,HinCop97,Hin02},
which predicts not only the linear scaling law for
domain walls, but also the amplitude of the relation. Clearly, a logarithmic
correction would be a problem for the approach. The main purpose of this paper
is to give a more accurate numerical determination of the scaling law (in 2
and 3 dimensions) and to determine the amplitude, to check the accuracy of the
theoretical predictions in \cite{Hindmarsh:1996xv,Hin02}. 

Our results for the scaling exponents and amplitude can be found
in Tables \ref{t:results2d}, \ref{t:results3d}. They are consistent with the 
linear scaling law for $ \xi  $, but the numerically determined
amplitudes are higher than the theoretical predictions. A detailed comparison 
can be found in \cite{Hin02}. 

\section{Dynamics of domain walls}

Domain walls occur in field theories whose manifold of minimum
energy states is topologically disconnected (see e.g.\ \cite{VilShe94}). The
canonical example in relativistic field theory is a theory of a
single scalar field $\phi(x)$, with action
\begin{equation}
 S = \int d^4 x
\sqrt{-g}\left( \half \pa_\mu \phi \pa^\mu \phi - V(\phi) \right),
\end{equation}
where the potential $V$ is a function with more than one
minimum, which we take to be the renormalisable form
\begin{equation}
V(\phi) = \frac{\lambda}{4}(\phi^2 - \mu^2)^2.
\end{equation}
In the cosmological context, the metric $g_{\mu\nu}$ is taken to
have the Friedmann-Robertson-Walker form
\[
g_{\mu\nu} = R^2(t)\eta_{\mu\nu},
\]
where $\eta_{\mu\nu} = \mathrm{diag}(1,-1,-1,-1)$ is the Minkowksi space 
metric and $t$ is conformal time.
The field can be conformally rescaled $R(t)\phi \to \phi$, giving
an Euler-Lagrange equation 
\begin{equation}
\label{e:LangEqMot} \ddot\phi + 2\frac{\dot R}{R}\dot\phi
-\nabla^{2}\phi +\lambda \phi \left( \phi ^{2}-\mu^{2}R^2\right) =0.
\end{equation}
In the broken phase, ($\mu^2 > 0$) there are domain wall solutions in which
the field changes vacuum over a distance of order $M^{-1}$, passing through
zero.  The solution for an infinite static planar wall is well known
\cite{VilShe94} and with an appropriate choice of coordinates can be written
\begin{equation}
\label{e:DWsoln}
\phi = \mu \tanh(Mz).
\end{equation}
 At high temperature the mass parameter receives thermal
corrections from the fluctuations in fields to which $\phi$ is
coupled \cite{Kirzhnits:1972iw,Kirzhnits:1976ts}:
\[
\mu^2(T) = \mu_0^2 - cT^2,
\]
where $c$ is a model-dependent numerical factor.  For the pure scalar field
theory $c=1/6$.  Thus as the Universe cools the field undergoes a phase
transition from $\vev{\phi} = 0$ to $\vev{\phi} = \pm\mu(T)$.  As this
transition happens at a finite rate, the field cannot select the same minimum
everywhere at the same time, and the Universe divides into domains in which
$\phi$ takes either positive or negative values. By continuity of the
field, these domains must be separated by domain walls \cite{Kibble:1976sj} in
which the field approximates the configuration given by Eq.\ \ref{e:DWsoln} in 
the transverse
direction. The initial size of the domains $\hat\xi$ is controlled by the rate
at which the transition occurs and how strongly damped the
field is \cite{Zurek:1996sj}.

The domain walls are mostly in the form of one infinite boundary
separating percolating clusters of the two vacua 
\cite{Ryden:1989vj,Coulson:1996nv,Larsson:1997sp}.
The subsequent evolution of the field is controlled by the
dynamics of this infinite domain wall.  The wall has a tension of
order $M^3$, and tries to straighten out and lose energy, and in
finite volume eventually one or other vacuum will take over the
whole space and the field thereby reaches equilibrium. One way of
quantifying the approach to equilibrium is to measure the area
density of the domain wall $\A$, or equivalently the curvature
scale of the wall $\xi = 1/\A$, where \cite{Hindmarsh:1996xv}
\begin{equation}
\label{e:AreaCountMethod}
A= \left\langle \delta(\phi) \left|
\nabla \phi \right| \right\rangle .
\end{equation}
At late times, this quantity can only depend on time $t$ and
the mass scale $M$.  The fact that the wall obeys a Nambu-Goto
equation independently of $M$ \cite{VilShe94} indicates that $\A =
a/t$ purely on dimensional grounds, where $a$ is a constant amplitude. 
This can be called the naive or
``classical''
scaling hypothesis for domain walls, which has been put on a more
rigorous footing in \cite{Hindmarsh:1996xv,Hin02}. The scaling
hypothesis, and the theory of dynamic scaling, can be also be
applied to other extended topological defects such as cosmic
strings.  More generally, one can define a scaling exponent $b$,
such that $\A = a/t^b$, and one of the goals of this paper is to measure
both the exponent $b$ and the amplitude $a$ as accurately as possible.

The first numerical simulations of this system were performed 
by Press, Ryden and
Spergel \cite{Press:1989yh,Ryden:1989vj} 
who noted that in comoving coordinates the width of the wall shrinks 
as $R^{-1}$, and so any numerical simulation with a lattice spacing 
fixed in comoving coordinates runs the risk of failing to resolve 
the domain wall at late times.  They showed that ignoring the $R$ 
dependence of the $\mu^2$ parameter did not substantially affect the 
dynamics of the walls, and we adopt the same approach.  We also include 
a damping term to simulate cooling in the early stages of the evolution.
Written in first order form, the field equations become 
%
\begin{eqnarray}
\label{e:HamEqMot}
\dot{\phi } & = & \pi \label{e:HamEq1} \\
\dot{\pi } & = & \nabla ^{2}\phi -\lambda \phi
\left( \phi ^{2}-\mu ^{2}\right) -\left(\frac{\dot R}{R}+\eta\right)\pi, 
\label{e:HamEq2}
\end{eqnarray}
The equations may be rescaled
to $ \mu =1$, $ \lambda =1$, so that the width of the wall is 1. 
These classical equations can be thought of as effective equations
representing the long-wavelength dynamics of the quantum field
when the occupation number is high.

The results of \cite{Press:1989yh,Ryden:1989vj} 
seemed to show that a
good fit for the area scaling law was given by
\begin{equation}
\label{e:logplaw}
A \propto \ln (t)/t,
\end{equation}
in both 2 and 3 dimensions, which is not predicted by the theory
of dynamic scaling for domain walls \cite{Hindmarsh:1996xv,Hin02}.  It
is therefore important to check these numerical simulations, and
after a decade of development in computer technology one can do
much larger simulations in order to eliminate transient effects.
Indeed, our largest simulation is performed on 3D grid of
$4096^3$, which gives us a dynamic range of roughly three orders
of magnitude between $t$ and $M^{-1}$, the characteristic response
time of the field \footnote{Such large simulations are possible only
by having part of the simulation volume in memory at any one time, and
keeping the rest on disk.}.

\section{Numerical Simulations}
\label{s:NumSim}

\subsection{Evolution Algorithm}

We used 2 and 3D cubic lattices with a 2nd order discretization of the
Laplacian operator.  The evolution of the discretized system was effected with
the Verlet or leapfrog algorithm, common 
in molecular
dynamics and offering a simple but effective way of discretizing 
the equations of motion (\ref{e:HamEq2}). 
The  scheme is
\begin{equation}
\ba{rcl}
\pi_{n+\frac{1}{2}} & = & \pi_{n-\frac{1}{2}}+f\left( t,\phi_{n},\pi_{n}\right)
\Delta t\\
\phi _{n+1} & = & \phi _{n}+\pi_{n+\frac{1}{2}}\Delta t,
\ea
\end{equation}
where $f\left( t,\phi _{n},\pi_{n}\right)  $ is the right hand side of
Eq.\ (\ref{e:HamEq2}), and $\pi_n = (\pi_{n+\frac{1}{2}} -
\pi_{n\frac{1}{2}})/2$.  As $f$ is linear in $\pi_n$, the equation can 
easily be rearranged so that $\pi_{n+\frac{1}{2}}$ is on the left hand side.
 The simulations on which the data is 
based all have $\Delta x =0.3$ and $\Delta t =0.1$.

We chose periodic boundary conditions, and restricted the
length of the simulation to a maximum time equal to $
T_{tot}=N\Delta x/2\Delta t, $ the time required for two signals
emitted from the same point and travelling in opposite directions
to interfere with each other.  This represents the time it takes for the field 
to ``notice'' the finite dimension of the lattice it resides in.

\subsection{Initial Conditions}

\begin{figure}
\begin{center}
\includegraphics[height=2.6 in,keepaspectratio=true]{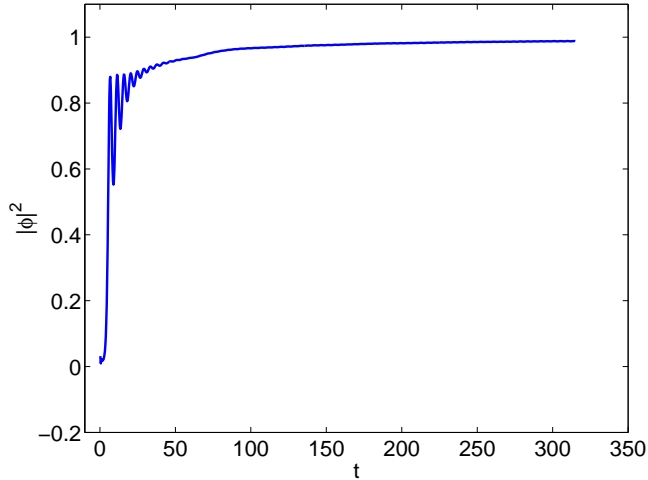}
\caption{Typical evolution of $ \left| \phi \right| ^{2} $}
\label{fig:ModPhi}
\end{center}
\end{figure}

The main objective of the simulations was not to see the formation of the 
network, but the evolution at later times, and so it is not necessary to 
start with a proper thermal distribution.  Instead, we try to imitate the 
configuration of the field in the potential
at some temperature above the critical $ T>T_{C} $, with a
Gaussian distribution around zero in $ k $-space. We created a
Gaussian distribution by adding 12 uniformly distributed random
numbers of unit variance, and algorithm that is quick and easy to
parallelize. By setting $ \phi _{k=0}=0 $ we set the average of
the distribution to zero and the proceed with transforming the
field to the $ x $-representation. This is an alternative to
simulating the actual phase transition by deforming the
symmetry-breaking potential. The Gaussian
distribution was chosen to give a pointwise spatial variance of about 
$10^{-1}\mu$, and the value of the volume-averaged field was 
always less than 10$ ^{-8}. $ Bigger ranges for the
Gaussian distribution were found to lead to instabilities 
unless high dissipation was used. This is due to the occasional large 
values of the field, which have high energy densities due to the quartic 
potential. After
initialisation the field is left to roll slowly to the two minima
and start oscillating in them. A typical evolution of $ \left|
\phi \right| ^{2} $ can be seen in Fig.\ \ref{fig:ModPhi}.

If the average field were not close to zero in the initial conditions 
distribute we would have found that the biased initial conditions would give a 
different time dependence in the area density 
\cite{Coulson:1996nv,Larsson:1997sp}. Indeed, if one phase is selected in 
the initial distribution then the evolution of the domain wall area
density can be shown to follow the relation
\begin{equation}
\label{e:PowerLawBiased}
A \propto \frac{1}{t}e^{-ct},
\end{equation}
where $ c $ a positive real number. Percolation theory predicts that for
a phase occupying a fraction less than a critical value $ p_{c}, $ the
infinite wall disappears. The dominant phase takes over at a
characteristic time scale, and quickly fills the entire simulation
volume. In our case $ p \simeq 0.5\pm 10^{-7}, $ far above the
percolation threshold which for a cubic lattice can be shown to be
$ p_{c}=0.31. $

\subsection{Beginning the Evolution}

In the beginning of the simulation dissipation is imposed in order for the 
field to sink in a controlled
manner into the minima and to remove spurious high frequency
modes. This is controlled by the parameter $\eta$ in
(\ref{e:HamEq2}).
 The dissipation aids the formation of the domain wall network and
as soon as this is formed, dissipation is turned off and the
system is left to evolve freely. A small amount of time, roughly
the system's characteristic time, is required for the field to
adjust itself into the new equation with no dissipation and
continue to evolve. For that reason any regression on data starts
a few timesteps after the end of the dissipated period. Turning
the dissipation off slowly seems to help the system adjust more
quickly to the new conditions and smooths out the effects of this
`adjustment' period. More specifically, the end of the
dissipation period $ T_{D} $ and the beginning of the regression $
T_{R} $ are specified in the program and their difference $
T_{R}-T_{D} $ is calculated. As soon as the system reaches $
T_{D}, $ dissipation at timestep $n$ $\eta_n$ is
decreased till zero using a Gaussian like
function
\begin{equation}
\eta_n = \eta_{\rm i} \times \exp \left( \frac{-A\left(
n-T_{D}\right) ^{2}}{\left( T_{R}-T_{D}\right) ^{2}}\right) \times
\Theta(T_{R}-T_{D}) ,
\end{equation}
where $ n $ is the timestep, $\eta_{\rm i}$ the initial
dissipation and $A$ a real positive number controlling in more
accuracy the length of this period.

Dissipations ranged from 0.1 to 0.3 in the simulations and
affected the system for roughly 10\%-30\% of the total simulation
time $ T_{tot}, $ with regression on the results starting at
20\%-40\% of the total time and the coefficient $ A $ was taken
to be in the range $ 1 $ to $ 1.5 $ .

\subsection{Wall Area Density Calculation}

The main purpose of the simulations was to check the power law for
the evolution of the wall area density. A crude way of calculating
the total wall area is counting the places where adjacent lattice
points have opposite signs and multiplying by $\Delta x$ or $ \Delta x^{2} $
which is a rough estimation of the wall area at the ``link''. More
precisely, one should find an accurate discretization of the
continuum area density operator (\ref{e:AreaCountMethod}).

The calculation is made by finding two adjacent lattice points
where the field has opposite signs and calculating the gradient of
the field at the ``link''. For two such adjacent points $ \left\{
i,j,k\right\}  $ and $ \left\{ i-1,j,k\right\}  $ the gradient is
calculated as follows. The numerical approximation to the gradient
at the direction of the link is just
\[
\Delta_{i}\phi =\frac{\phi_{i,j,k}-\phi _{i-1,j,k}}{\Delta x}
\]
 whereas the gradients at the remaining two directions are taken by averaging
over the gradients of nearby links; the $ j $ -component of the
gradient for example would be
\[
\Delta_{j}\phi =\frac{1}{2}\left( \frac{\phi _{i-1,j+1,k}-
\phi_{i-1,j-1,k}}{2\Delta x}+\frac{\phi _{i,j+1,k}-\phi
_{i,j-1,k}}{2\Delta x}\right) .
\]
One needs to account for the orientation of the area element at
each link, otherwise one will end up overestimating the area
\cite{Ryden:1989vj,Scherrer:1998sq}. For domain walls in 3D the
problem can be resolved by simply multiplying the lattice area
estimate by a factor $ \frac{2}{3} $ \cite{Scherrer:1998sq}, while 
a similar argument to that given in \cite{Scherrer:1998sq} gives the factor 
$\pi/4$ in 2D: 
\bea
A &=& A_{Lat} \frac{2}{3}
\qquad \textrm{(3D)} \\
A &=& A_{Lat} \frac{\pi}{4}
\qquad \textrm{(2D)}
\eea
The Figures show $A_{Lat}$, while the Tables show 
$A$.

\section{Results}

\subsection{Wall area density}

\begin{figure}[t!]
\begin{center}
\includegraphics[height=2.6 in,keepaspectratio=true]{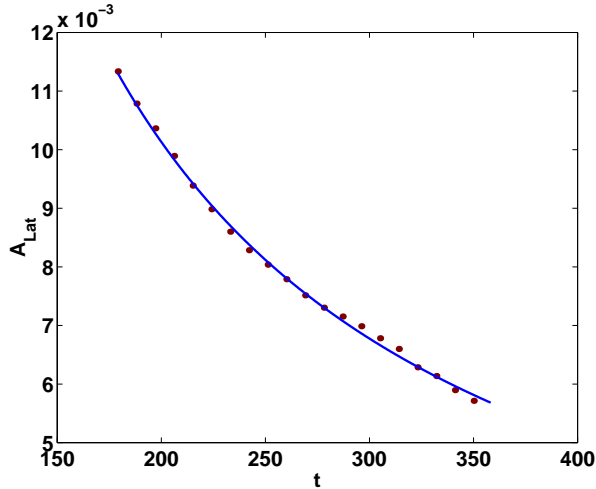}
\caption{Evolution of the wall area density $A_{Lat}$ for a 2D
simulation, $N=1024$, $\Delta x =0.3$, $\Delta t =0.1$. } \label{fig:Area}
\end{center}
\end{figure}

In all simulations the wall area data 
started to be taken shortly after the dissipation had been stopped.
Fig.\ \ref{fig:Area} shows the results from a 2D run, fitted to a 
power law 
\begin{equation}
\label{e:powerlaw}
A_{Lat}=at^{-b}
\end{equation}
 by regression. Five simulations 
for the same parameters were run and an
average was taken, with the fit taken on the averaged area.
The results for 2D and 3D simulations are
presented in Tables \ref{t:results2d}, \ref{t:results3d}, for Minkowski, 
\begin{table}[ht]
\begin{tabular}{|l|l|}
\hline
\textit{Background}
        & \textit{Length scaling law} \\
\hline
%
%
Minkowski & $0.765(0.227).t^{-0.987(0.032)}$
\\
Radiation & $0.928(0.165).t^{-0.996(0.018)}$
\\
Matter    & $1.145(0.227).t^{-0.992(0.014)}$  \\
\hline
\end{tabular}
\caption{\label{t:results2d}
Area scaling laws for Minkowski,
radiation-dominated, and matter-dominated FRW
backgrounds in 2 dimensions. The results are derived from averaging over 5
simulations for each case, with ${\Delta x}=0.3$, ${\Delta t} =
{0.1}$ and a $1024^2$ lattice.}
\end{table}
\begin{table}[ht]
\begin{tabular}{|l|c|}
\hline
\textit{Background}
        & \textit{Area scaling law}  \\
\hline
%
%
Minkowski
& $0.883(0.141). t^{-0.995(0.026)}$  \\
Radiation
& $0.925(0.125). t^{-0.994(0.013)}$  \\
Matter    & $0.963(0.122). t^{-0.997(0.012)}$  \\
\hline
\end{tabular}
\caption{\label{t:results3d}
Area scaling laws for Minkowski,
radiation-dominated, and matter-dominated FRW
backgrounds in 3 dimensions. The results are derived from averaging over 5
simulations for each case, with ${\Delta x}=0.3$, ${\Delta t} =
{0.1}$ and a $512^3$ lattice.}
\end{table}
radiation and
matter-dominated Friedmann-Robertson-Walker backgrounds
respectively. 
The biggest 3D simulation ($ N=4096 $, $\Delta x =0.3$, $\Delta t = 0.08$)
gave  
$ a=0.980 ( \pm 0.017 )$,  
$ b=0.985 (\pm 0.003 ).$

\begin{figure}[ht]
\begin{center}
\includegraphics[height=2.6 in,keepaspectratio=true]{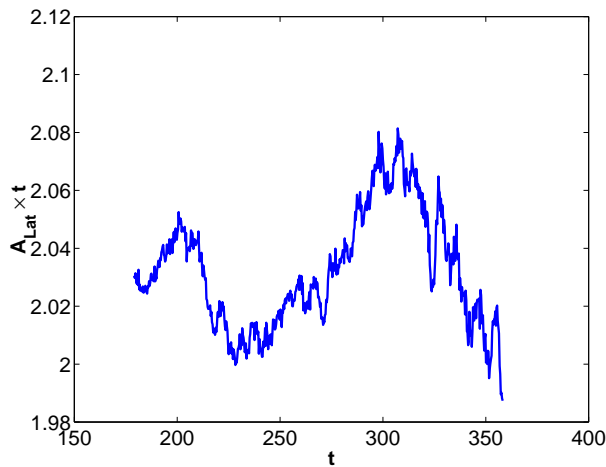}
\caption{ $\exp(A_{Lat}t) $ against time for the simulation of Fig.
\ref{fig:Area}.  A decay law $A_{Lat}\sim \log(t)/t$ would 
show as a logarithmic increase on this graph.} \label{fig:NoLog}
\end{center}
\end{figure}

There is a suggestion from the simulations 
that the wall area decreases slightly
slower than $ b=1. $ However, this deviation from the $ b=1 $
scaling could not be attributed to a relation of the form
given in Eq.\ \ref{e:logplaw}, as it has been suggested \cite{Press:1989yh}.
Plotting $ \exp(A_{Lat}) \times t $ against time shows that
there seems to be no logarithmic term in the wall area evolution,
Fig. \ref{fig:NoLog}.

Both the power law and the coefficient $ a $ of Eq.\  \ref{e:powerlaw}.
present challenges to the analytic method for the calculation of the wall area
density of Ref.\ \cite{Hindmarsh:1996xv}, which 
are compared in detail in \cite{Hin02}.
The power law is very close to the predicted value of $b=1$, 
with good precision, but the coefficient $ a $
showed larger 
fluctuations between runs.

\section{Conclusions}

The numerical integration of the equations of motion for the the
$\phi^4$ model
has given an insight to the dynamics involved in domain wall networks and has
provided an accurate way to support the scaling solution predicted by 
theoretical
computations. A power law with exponent very close to 1 was found to be the
best solution according to the simulation data with no evidence for a
logarithmic term suggested in previous studies \cite{Press:1989yh}.

\begin{figure}[ht!]
{\centering
\begin{tabular}{cc}
 {\scalebox{.41}{\includegraphics{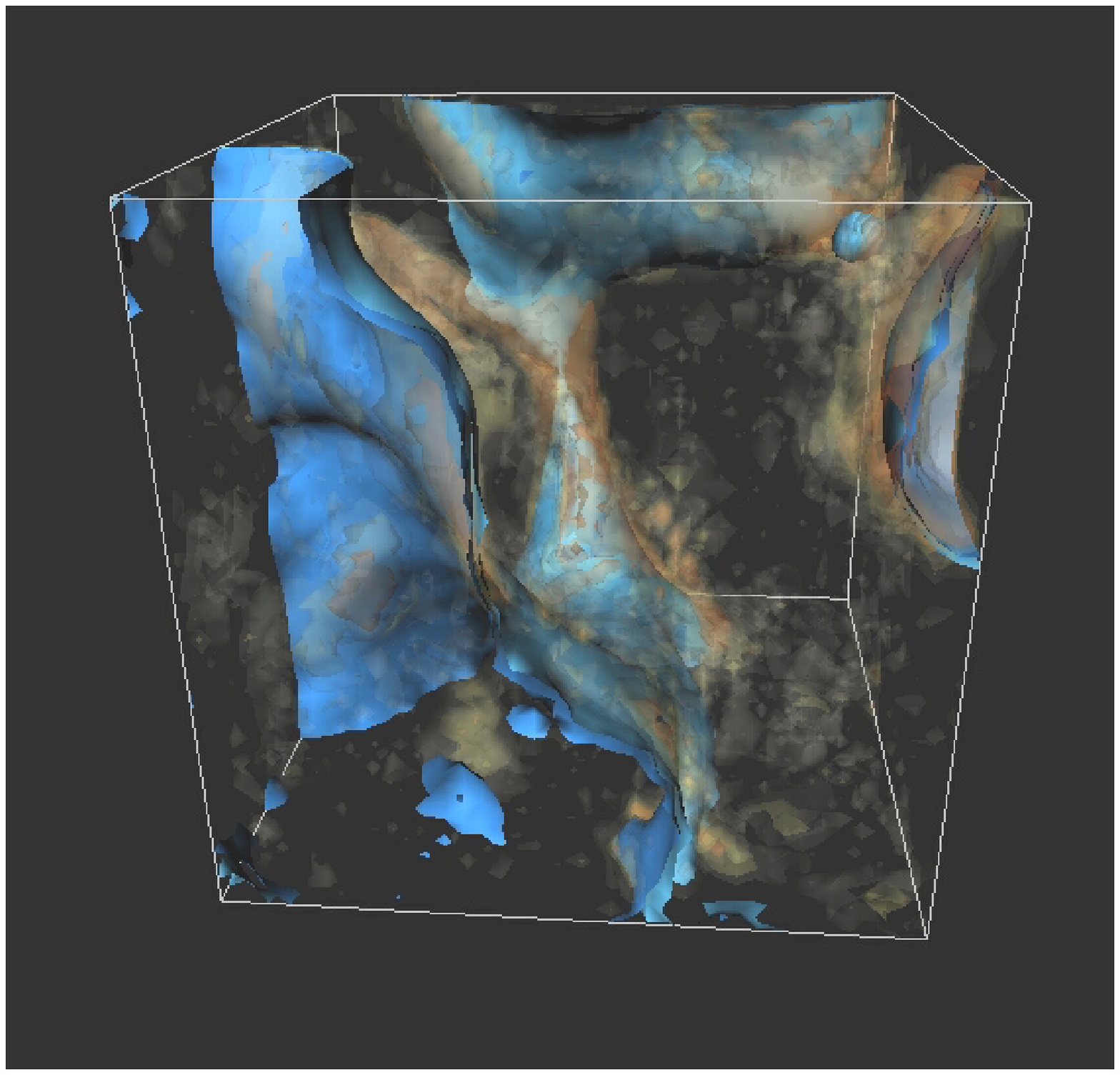}}} &
 {\scalebox{.41}{\includegraphics{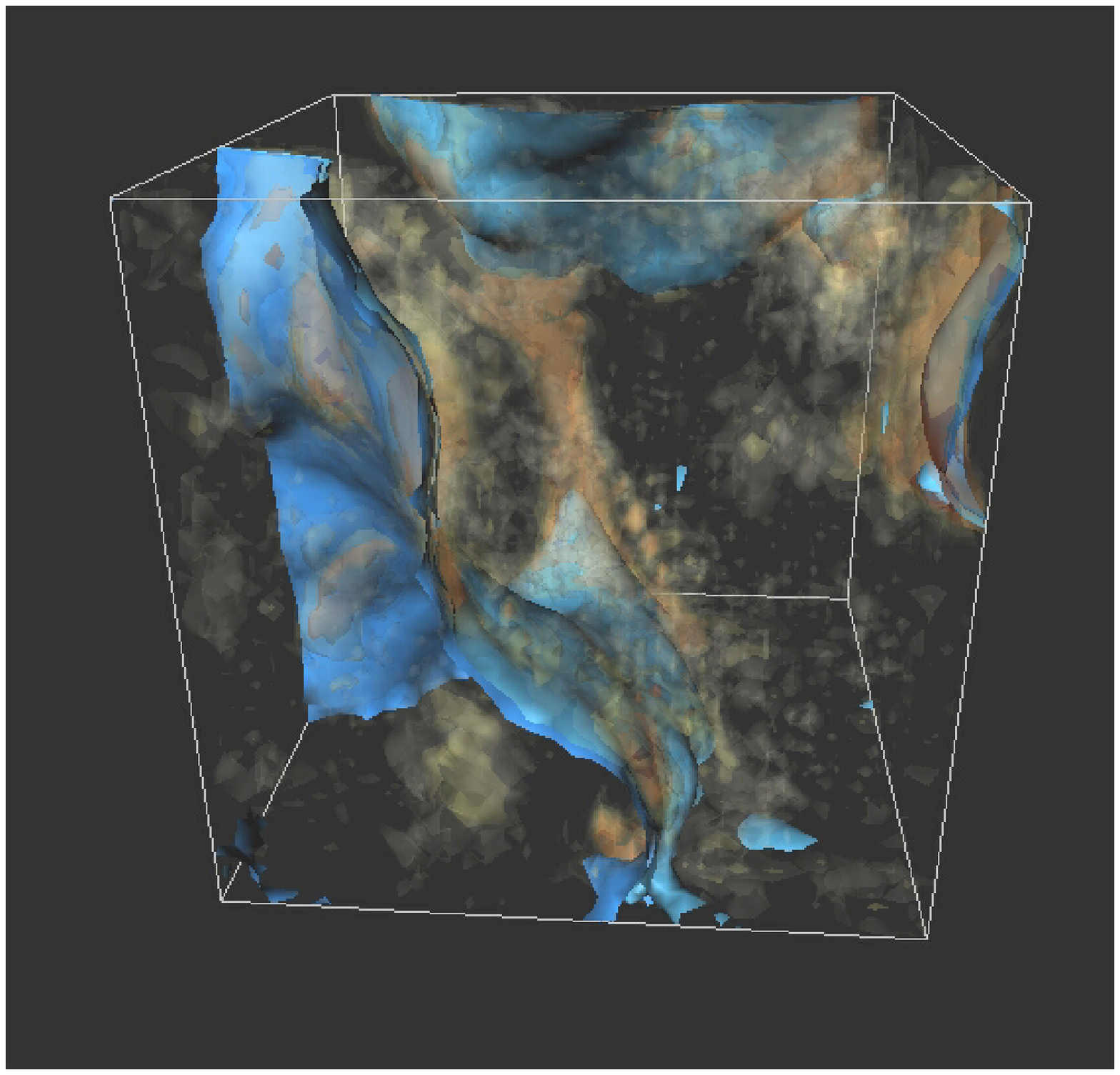}}}\\
 {\scalebox{.41}{\includegraphics{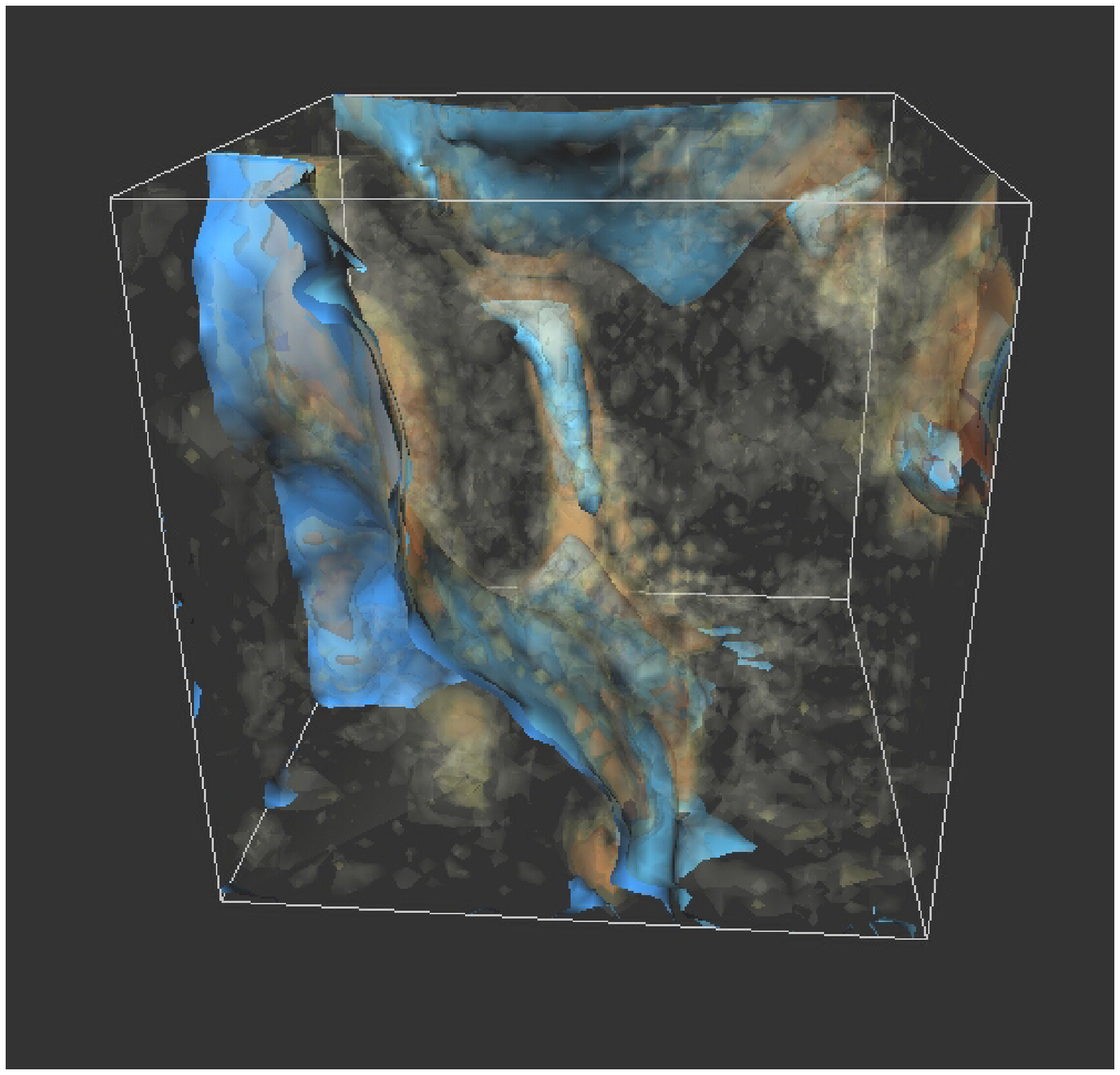}}} & 
 {\scalebox{.41}{\includegraphics{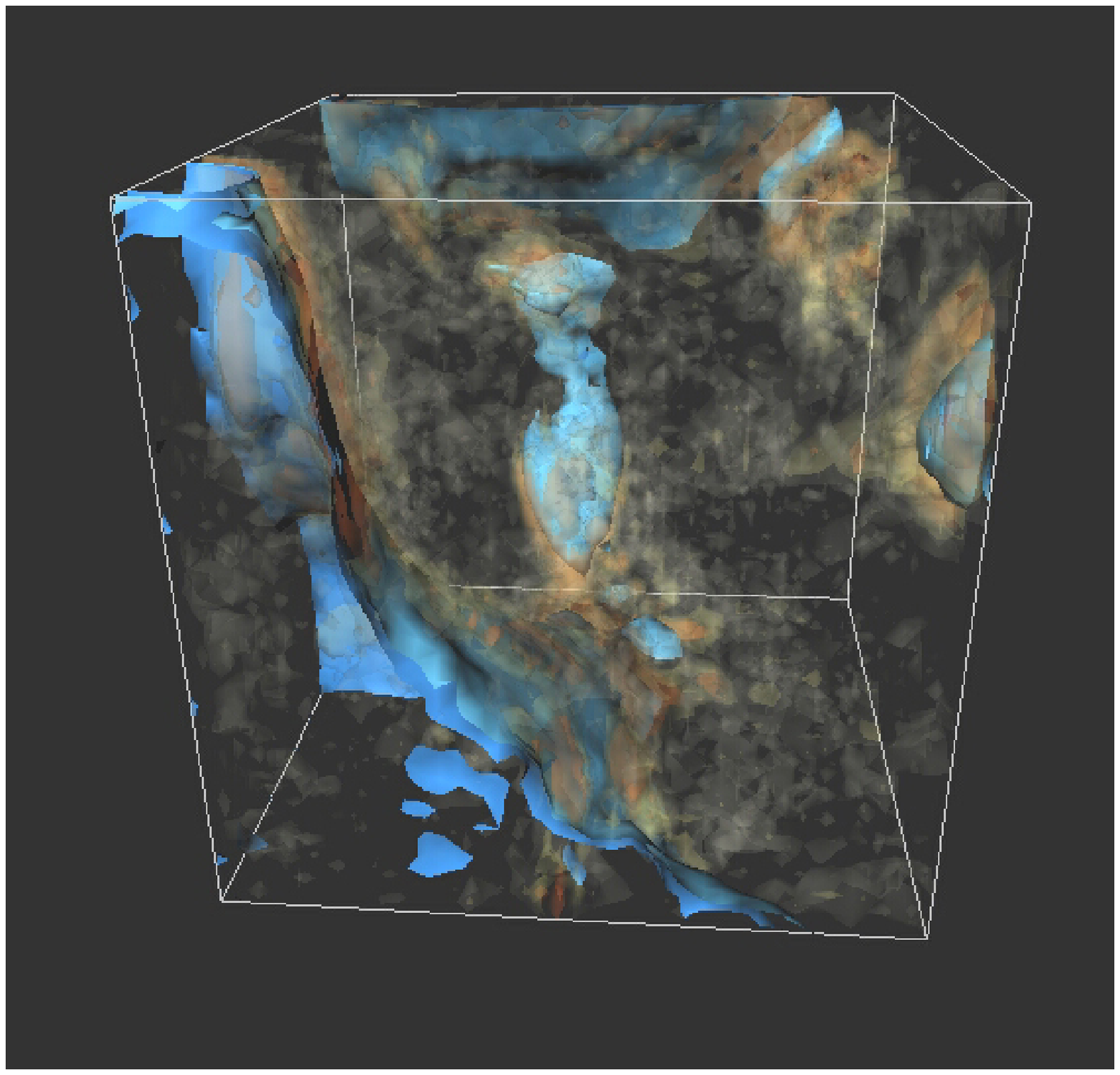}}}\\
 {\scalebox{.41}{\includegraphics{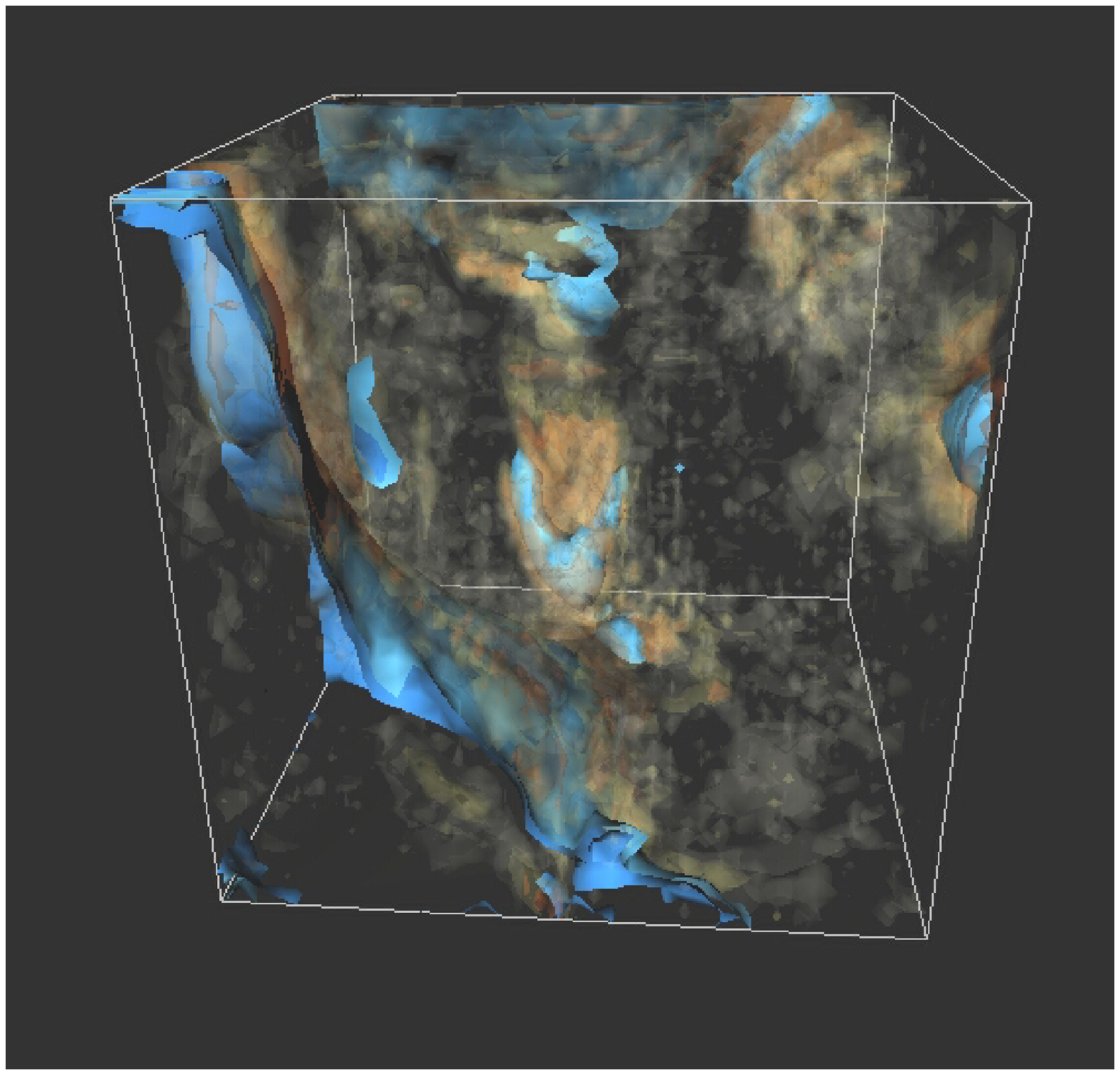}}} &
 {\scalebox{.41}{\includegraphics{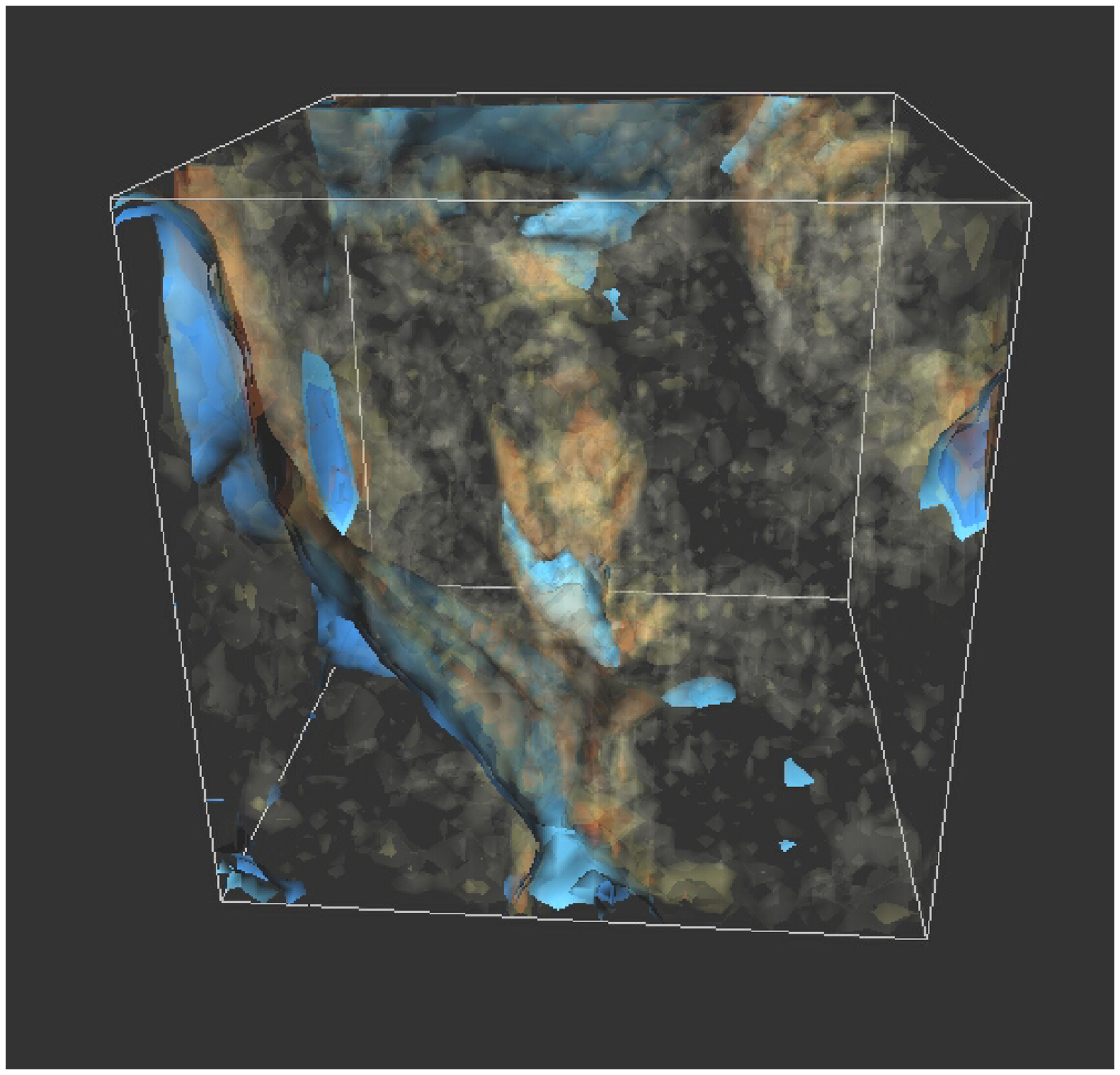}}}
\end{tabular}
}
\caption{\label{f:walls}
Six snapshots from a numerical simulation using the algorithm described 
in Sec.\ \ref{s:NumSim}, with time increasing left to right and top to bottom.
The solid isosurfaces are surfaces of constant
$\phi$, with the red representing small positive values and the blue small 
negative values. 
Semi-transparent isosurfaces are surfaces of constant momentum density.}
\end{figure}

The fact that a domain wall network shows this dynamic scaling
over approximately three orders of magnitude in the parameter
$M\xi$, the ratio between the local curvature radius and the wall
width, is remarkable, and has important implications for cosmic
string networks. Firstly, it is clear from the visualisations
in Fig.\ \ref{f:walls} (see also \cite{Sims})
that the energy in the domain walls is very quickly
transferred into propagating modes of the field: the formation and
collapse of closed loops (2D) or surfaces (3D), expected in the
standard picture of the evolution of wall networks
\cite{VilShe94}, is rare.  Indeed, in 2D, self-intersections to form
closed loops must be very rare; if two segments of wall approach
each other they must be generically curved away from the point of
closest approach, and therefore the acceleration is in the
direction which would tend to increase the separation.  Nonetheless, 2D
walls scale perfectly well, so it seems plausible in that
case that energy is being transferred {\it directly} into
radiation.  If the amplitude of the oscillations is large enough they 
could appear to form tiny loops or ``protoloops''
\cite{Moore:1998gp,Moore:2001px}.

We believe that our results add weight to the contention, first
put forward in \cite{Vincent:1998cx}, that extended defects
(including cosmic strings in 3D) have a non-perturbative channel
into massive radiation. At first sight this is difficult to square
with the standard picture, in which walls and strings obey the
Nambu-Goto equations of motion for large curvature radii, for in
that case the total energy locked up in the extended defects is
conserved, in the absence of an general relativistic effects such
as an expanding background or gravitational radiation.  It is
certainly true that it is possible to find string trajectories
which are very close to being solutions of the Nambu-Goto
equations \cite{Moore:1998gp,Olum:2000sg}: however, the initial
conditions have to be carefully prepared, and the existence of
these trajectories does not preclude the existence of a
non-perturbative radiative process for defect networks.  Indeed,
we maintain that our results are good evidence that there must be
such a process.

\acknowledgments
This work was conducted on the SGI Origin platform using COSMOS
Consortium facilities, funded by HEFCE, PPARC and SGI.
We also acknowledge support from the Sussex High Performance Computing
Initiative.

\appendix

\section{Lattice Correction factor for 2D wall length density}

In Ref.\ \cite{Scherrer:1998sq} it was shown that the naive lattice estimate 
of the area density of random domain walls in 3D overcounts the continuum
value by a factor of 3/2 in the limit that the radii of curvature are large
compared with the lattice spacing.
 In this section we perform the analagous calculation
for two dimensions, finding it to be $4/\pi$.

The naive estimate is obtained by summing the
length of all links containing a wall and dividing by the total
volume.  A link crossing a wall is defined to be one for which the values of
the field $\phi$ on the sites at either end have opposite signs.  One can
immediately see this will overestimate the length, as one is approximating a
smooth curve by a sequence of line segments parallel to the lattice vectors
$\bfi$ and $\bj$.

In the continuum the centre of a domain wall is described by 
a 2-dimensional curve $\br(\si)$. Consider a segment of length
$l$, with $l \ll \xi$, where $\xi$ is the correlation length of the curve.
Writing $\De\br = \br(l)
- \br(0) = x\bfi + y \bj$, we see that the lattice approximation
to the length is
\ben
l_{\mathrm{Lat}} = |x| + |y|.
\een
In the limit that $l/\xi \to 0$, the continuum value of the length
is $l= \surd(x^2+y^2)$. Hence 
\ben
l_{\mathrm{Lat}} = \frac{l_{\mathrm{Lat}}}{l}l  = (|\cos\al| + |\sin\be|)l
\een
where $\al$ and $\be = \pi/2 - \al$ are the direction cosines of the vector 
$\De \br$. The ratio between the lattice estimate of the length and the true
length is obtained by averaging over all possible orientations of the lattice
relative to $\De \br$:
\ben
\left\langle\frac{l_{\mathrm{Lat}}}{l}\right\rangle = 
\frac{1}{2\pi}\int_0^{2\pi} d\al \left( |\cos\al| + |\sin\al|\right) =
\frac{4}{\pi},
\een
as advertised.

This calculation can be extended to arbitrary $l$ through a more involved
argument.  Let us first define the correlation function
\ben
C(r) = \vev{\phi(\br)\phi(0)},
\een
which is assumed to be smooth at $\br=0$, so that 
\ben
C(r) = C(0) - \half C''(0)r^2.
\een
In Refs.\ \cite{OhtJasKaw82,Hindmarsh:1996xv}
it is shown that the length density $A$ 
of the locus of zeroes of a
Gaussian random field in 2 dimensions is given by 
\ben
A = \frac{1}{2}\sqrt{-\frac{C''(0)}{C(0)}}.
\label{e:contl}
\een
This is the value of the length density in the continuum.

On the lattice, we must consider the probability that the values of the field
at opposite ends of a link have
opposite signs, in which case we can say that the link
is occupied by a segment of domain wall. Let us call this probability
$p_{\mathrm{occ}}$, which is
\ben
p_{\mathrm{occ}} = 2P(\phi(\bx)> 0\; \mathrm{AND}\; \phi(\bx + \bfi\De
x) < 0),  
\een
where the factor of 2 accounts for the opposite case 
$\phi(\bx)> 0\; \mathrm{AND}\; \phi(\bx + \bfi\De
x) < 0$.  The lattice estimate 
of the length density is then
\ben
A_{\mathrm{Lat}} = \frac{2p_{\mathrm{occ}}}{\De x},
\een
where the lattice spacing is $\De x$, and the factor 2 comes from the fact that
there are twice as many links as sites in 2 dimensions.

Suppose we now define 
\ben 
P_{12}(C(r_{12}),V) = 
P(\phi(\bx_1)> V\; \mathrm{AND}\; \phi(\bx_2) < V),
\een
where $V$ is an arbitrary threshold and $r_{12} = |\bx_1 - \bx_2|$.
One can then almost trivially write
\ben
P_{12}(C(r_{12}),V) = \int_0^{r_{12}} \frac{\pa P_{12}}{\pa C(r)} \frac{d
C(r)}{dr} dr.
\een
It can be shown \cite{HamGotWei86} that 
\ben
\frac{\pa P_{12}}{\pa C_{12}} = \frac{1}{2\pi[C(0)^2 - C_{12}^2]^\half}
\exp\left( - \frac{V^2}{C(0) + C_{12}}\right),
\een
where $C_{12} = C(r_{12})$.  Hence
the lattice estimate of the area density is
\ben
A_{\mathrm{Lat}} = \frac{2}{\pi\De x}\cos^{-1}\left(\frac{C(\De
x)}{C(0)}\right).
\een
Providing $\De x$ is much smaller than the correlation length $\xi$, defined 
by $\xi^2 = |C(0)/C''(0)|$, we see that 
\ben
A_{\mathrm{Lat}} = \frac{2}{\pi} \sqrt{-\frac{C''(0)}{C(0)}}, 
\een
and hence, from Eq.\ (\ref{e:contl}), that $A_{\mathrm{Lat}} =4A/\pi$.

\bigskip

\end{document}